\RequirePackage{fix-cm}

\documentclass[smallextended]{svjour3}

\smartqed

\usepackage{color}

\usepackage{graphicx}
\usepackage[misc]{ifsym}

\usepackage{caption}
\usepackage{ulem}
\renewcommand{\figurename}{FIGURE}

\newcommand{\insight}{\textit{Insight}-HXMT}

\begin{document}

\title{The removal method and generation mechanism of spikes in \insight{}/HE telescope}

\author{Baiyang Wu\textsuperscript{1,2}    \and
        Yifei Zhang\textsuperscript{1}   \and
        Xiaobo Li\textsuperscript{1}     \and
        Haisheng Zhao\textsuperscript{1} \and
        Mingyu Ge\textsuperscript{1}     \and
        Congzhan Liu\textsuperscript{1}  \and
        Liming Song\textsuperscript{1}   \and
        Jinlu Qu\textsuperscript{1}}

\institute{\Letter \textcolor{black}{Xiaobo Li, Liming Song}    \at
          \email{\textcolor{black}{lixb@ihep.ac.cn, songlm@ihep.ac.cn}}
          \and
        }


\authorrunning{Baiyang Wu et al.}
\maketitle

\begin{abstract}
Spikes are some obvious sharp increases that appear on \textcolor{black}{the} raw light curves of \insight{}'s High Energy X-ray telescope(HE), which could have influences on data products like energy and power spectra. They are considered to be fake triggers generated by large signals. In this paper, we study the spikes' characteristic and propose two methods to remove spikes from the raw data. According to the different influences on energy and power spectra, the best parameters for removing the spikes is selected and used in the \insight{} data analysis software. The generation mechanism of spikes is also studied using the backup HE detectors on ground and the spikes can be reduced by the electronic design. \\
\end{abstract}

\section{Introduction}
\label{intro}

The Hard X-ray Modulation Telescope (HXMT), dubbed \insight{}, successfully launched on June 15, 2017\cite{Zhang_2020}\cite{li2020inflight}. It is China’s first X-ray astronomy satellite devoted to broad band observations in the 1--250\,keV. It carries three main telescopes: High Energy X-ray telescope\,(HE) using an array of NaI(Tl)/CsI(Na) scintillation detectors\cite{liu2019high}, covering the 20--250\,keV energy band, Medium Energy X-ray telescope\,(ME) using an array of Silicon Positive-Intrinsic-Negative (Si-PIN) detectors in the 5--30\,keV band \cite{cao2019medium}, and Low Energy X-ray detector\,(LE) using an array of Swept Charge Device (SCD) detectors in the 1--15 keV band\cite{chen2019low}.

We will give a short description of HE. HE utilizes 18 NaI(Tl)/CsI(Na) phoswich as the main detectors with diameter of 190\,mm. The thicknesses of NaI(Tl) and CsI(Na) are about 3.5\,mm and 40\,mm, respectively. 
The particle with most of its energy deposited in NaI(Tl) is regarded as a NaI(Tl) event. The scintillation photons generated in the two crystals can be collected by a shared photomultiplier tube (PMT). Signals from the PMT can be distinguished NaI(Tl) events and CsI(Na) events due to the different decay time in the two crystals. The energy loss, time of arrival and pulse width of each detected event are measured and digitized. The CsI(Na) can be used as a gamma-ray burst (GRB) monitor. The detected energy band in high gain mode is about 50--800\,keV and 250\,keV--3\,MeV in low gain mode for CsI(Na) if the high voltage of PMT is decreased\cite{li2020inflight}. The spikes in low gain mode can be ignored \textcolor{black}{because the large signals have been greatly reduced,  thus} we focus on the spikes happened in high gain mode.
 
There are similar spikes that have been found on other detectors or experiments which have provided some explanations to \textcolor{black}{the generation of} spikes. \textcolor{black}{The} hard x-ray detector of Astro-E appears fake pre-trigger signals which are due to after pulses of the photomultiplier\cite{10.1117/12.366546}. Photomultiplier tube would produce signal-induced noises that are usually considered to be the electron cloud that surrounds the photocathode or the fluorescence that is induced by ions striking the glass envelope of the PMT\cite{Zhang2018}. There are also more delayed after pulses that are possible results of the phosphorescence of glass or other insulators with activation energies well above kT\cite{2016RadM...86...39T}. In addition, space charge effects and statistical fluctuations in the background\cite{Meegan_2009} also contribute the spikes. 
 
Since HE detectors also consist of crystal components and PMTs, similar effects may cause spikes. According to our ground experiments (we describe details in section~\ref{sec:4}), we have found large signals would cause a series of effects that are mentioned above. These effects make large signals come out from the PMT with long tails which cause electronics continuously triggered and counted repeatedly, and finally spikes are appeared on the light curves. \\
 
Spikes could distort the shape of energy spectrum and power spectrum of the events. In order to obtain the spikes removed data products, we explore two methods to remove spike events and find out the optimal parameters in this paper.
We structure our paper as follows. In section~\ref{sec:2}, we describe how we use the observed data to study the spikes' features. In section~\ref{sec:3}, two methods of removing spikes are described and compared to find the best parameters for the data analysis of HXMT. The experiments setup is described to study the generation and reduction of spikes using the backup HE detectors on ground in section~\ref{sec:4}. In section~\ref{sec:5}, we give a conclusion and discussion.\\

\begin{table*}
    \centering
    \begin{tabular}{||c c c c c||}
    \hline
          Observation ID & Object & Observation start & Observation stop & Exposure(s) \\
    \hline\hline
         P0101299001 & Crab & 2017-08-27 04:05:32 & 2017-08-28 04:17:53 & 87141 \\
    \hline
         P0101297302 & Crab & 2017-09-13 08:15:05 & 2017-09-13 11:41:48 & 12403 \\
    \hline
         P0111605063 & Crab & 2018-08-30 07:05:01 & 2018-08-31 08:50:59 & 92758 \\
    \hline
         P0202041254 & Crab & 2019-12-04 05:11:06 & 2019-12-05 00:35:38 & 69872 \\
    \hline
         P0101293001 & Blank Sky & 2017-11-02 05:00:57 & 2017-11-02 17:56:12 & 46515 \\
    \hline 
    \end{tabular}
    \caption{The observation information}
    \label{tab:Crab}
\end{table*}
\begin{figure}
    \centering
    \includegraphics[width=\linewidth]{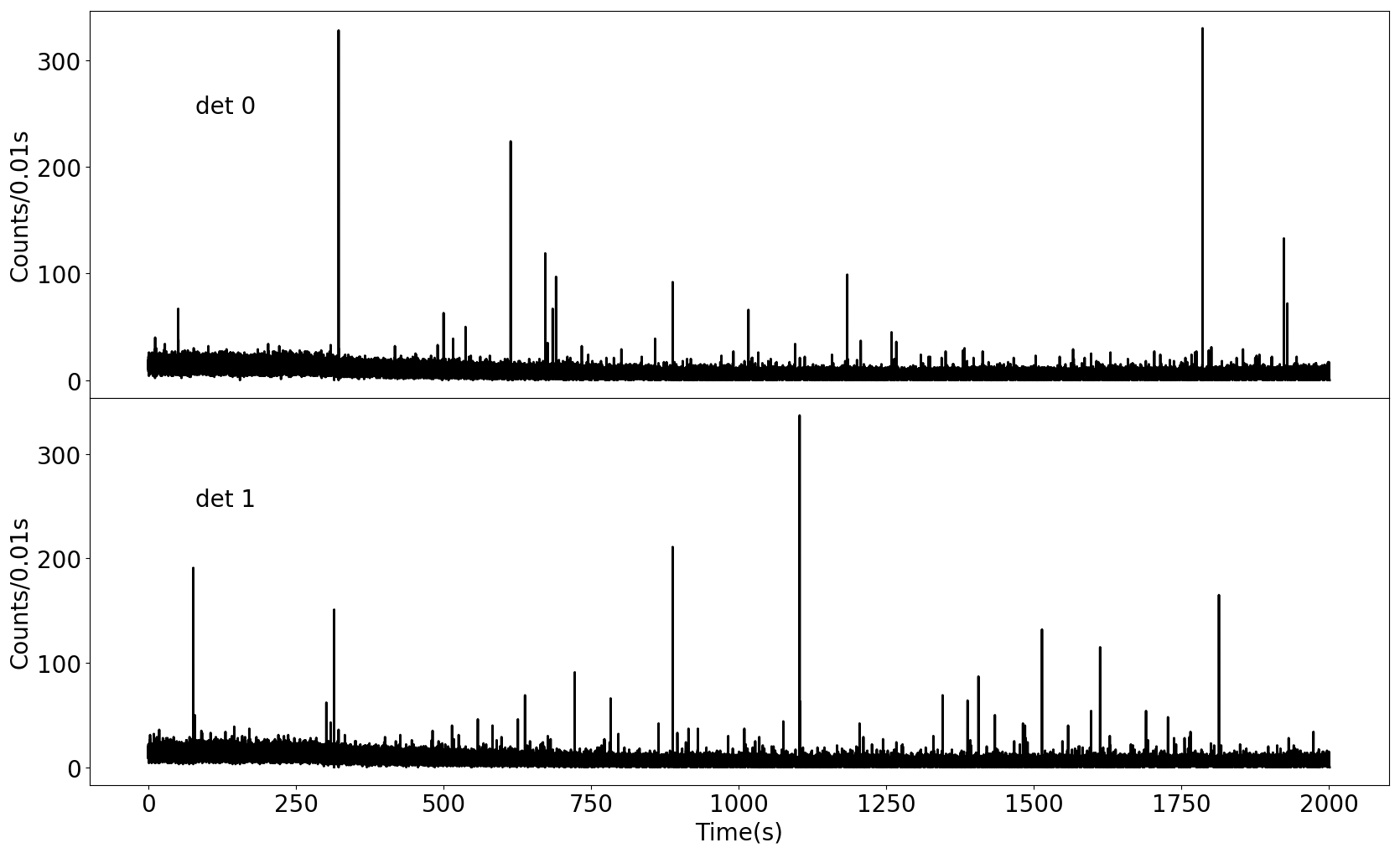}
    \caption{Raw light curves of detector ID 0 and 1, both with time resolution of 0.01\,s. There are some prominent spikes on both light curves.}
    \label{rawlc}
\end{figure}

\section{The features of HE spikes}
\label{sec:2}
Since the spikes are often appeared in the high gain mode of HE detectors, a lot of pointing observations of Crab and blank sky have been selected to analysis the features of spikes. The observation log of the data is summarized in table \ref{tab:Crab}\footnote{\textcolor{black}{All Insight-HXMT data used in this work are publicly available and can be downloaded from the official website of Insight-HXMT: http://archive.hxmt.cn/proposal.}}. 
As an example of spikes, a fragment of the raw light curve of one Crab observation with exposure time of about 2000 seconds is plotted in Figure \ref{rawlc}. The light curves of detector 0 and detector 1 are plotted respectively to show that the spikes occur in different detectors independently. We can see there are some apparent spikes whose counts are much larger than the statistic fluctuations on the raw light curve. \\

\begin{figure*}
    \centering
    \includegraphics[width=\linewidth]{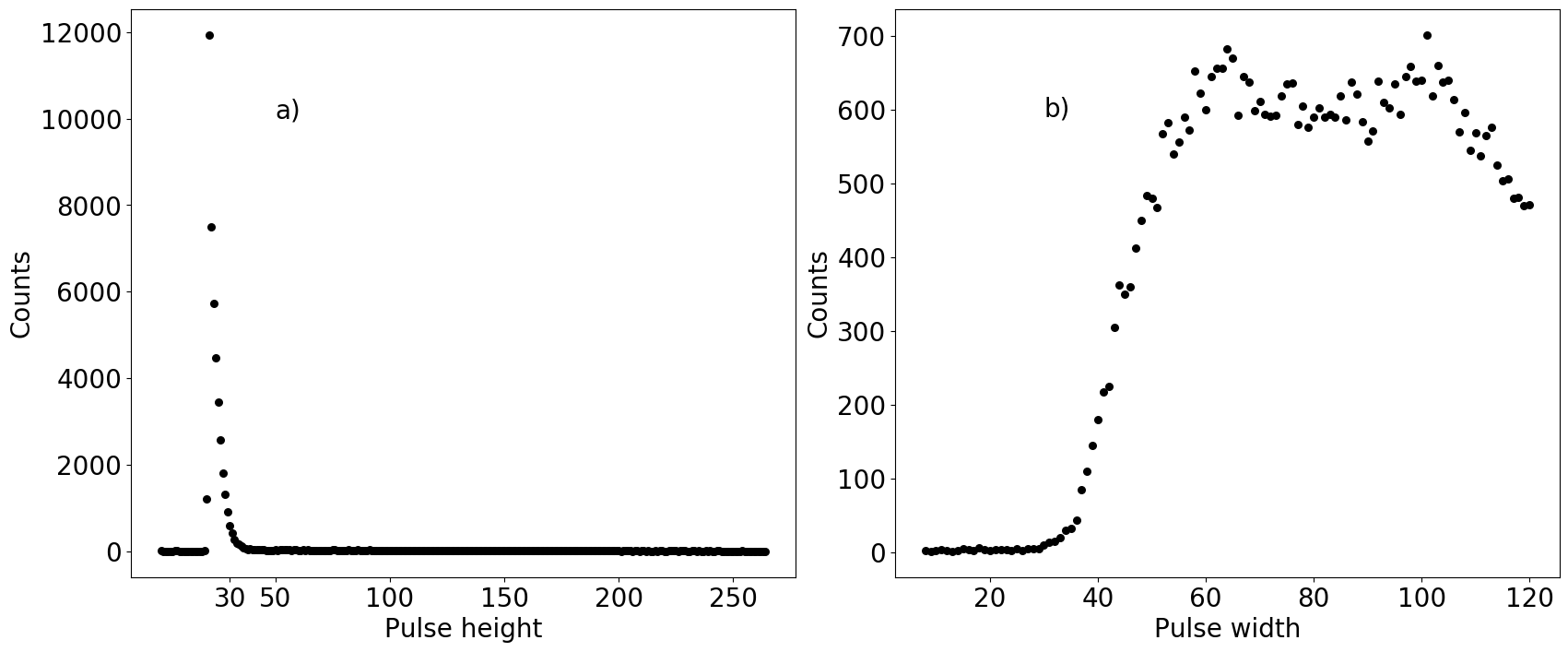}
    \caption{a)\textcolor{black}{Pulse height(PHA)} distribution of the prominent spikes. b)Pulse width distribution of the large spikes.}
    \label{FIG:1}
\end{figure*}

\begin{figure}[]
    \centering
    \includegraphics[width=\linewidth]{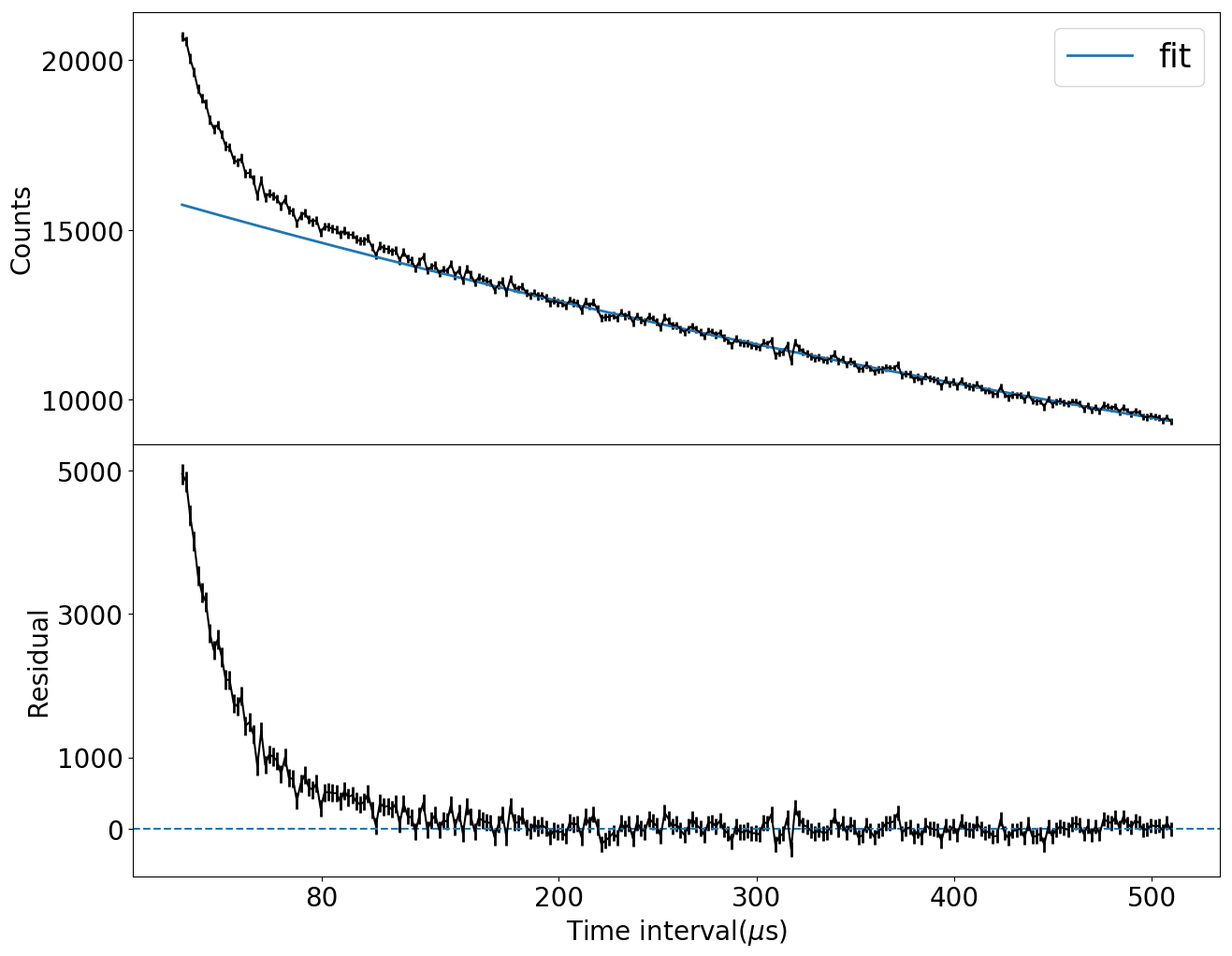}
    \caption{The top panel shows the distribution of time intervals between two adjacent events with unit of $\mu$s. The blue lines is the exponential function used to fit the distribution. The bottom panel shows the residuals of the fit result and data. The deviation from the exponential distribution caused by the spikes is less than 200\,$\mu$s.}
    \label{FIG:2}
\end{figure}

In order to explore the characteristic of the spikes, some sharp and prominent spikes are selected from the raw data. The distributions of pulse heights (PHA) and pulse widths of spikes are shown in Figure \ref{FIG:1}. It is clear that most of pulse heights of spikes are between the pulse height threshold\,(20) and 35 channel. Meanwhile, the distribution of pulse widths covers the range from 40 to 120 channel, where one channel means the pulse width of 20.8\,ms. Most of events with pulse widths under 70 channel are NaI events and the widths higher than 80 channel are CsI events. The distribution of pulse width indicates that spikes come from both NaI(Tl) and CsI(Na).\\
The distribution of time intervals between two adjacent events is shown in Figure \ref{FIG:2}. An obvious deviation from the exponential distribution can be found in the range of less than 200\,$\mu$s. We consider this deviation comes from the contributions of the spikes. \\
This characteristic of HE spikes can be summarized as:
 \begin{itemize}
  \item Most of the pulse heights for spikes is less than 35 channel.
  \item Pulse width covers the range of NaI(Tl) and CsI(Na) events.
  \item Most of the time intervals between two adjacent events are less than 200\,us.
\end{itemize}

\captionsetup[figure]{labelfont={color=black}}

\captionsetup[figure]{labelfont={color=black}}

\section{Methods of removing spikes and their influences}
\label{sec:3}
\subsection{Two methods to remove spike-like events}
\label{sec:3.1}
Spikes are some unusually dense events with shorter time intervals than normal events, we have come up with two methods to remove events that may be spikes. 
We mark a raw time series as $T_o$, which contains all events, marked as  $t_1,t_2,...t_i,t_{i+1},...,t_N,$ N is the total number of these events. \\

The first method to remove spikes is that if there is a number of consecutive events, 
$t_i,t_{i+1},...,t_{i+n}$, where\\
$t_{i+n}$-$t_i<t_f$ and $n+1 \geq n_f$,
we remove this consecutive events, i.e, we remove these n+1 events. $t_f$ and $n_f$ can be determined empirically. For simplicity, we call this method as method 1.\\

The second method\,(method 2) requires more than $n_s$ consecutive events where each time interval between two consecutive events is less than $t_s$, we remove these events that meet the conditions. $t_s$ and $n_s$ can also be determined empirically. \\

For each method to remove spike-like events, we basically follow these steps to analysis its effects. We choose a set of parameters, $t_f$, $n_f$ for the first method and $n_s$, $t_s$ for the second one. The events satisfied these specific parameters are regarded as the spike-like events and would be removed from the event file.\\

After removing the spike-like events, a new time series can be obtained as $T_r$. We analyse the new time series through three aspects: light curve, power spectrum and energy spectrum. For each method we have utilized several sets of parameters which are listed in Table \ref{table:1} and \ref{table:2}. The percentages of events that remain after removing "spikes" to the original events number are listed in the third column of Table \ref{table:1} and \ref{table:2}.\\
\begin{table}[]
\centering
 \begin{tabular}{||c c c c||} 
 \hline
   & $t_f$ & $n_f$ & percentage \\ [0.5ex] 
 \hline\hline
 1 & 1\,ms & 4 & 84.0\% \\
 2 & 1\,ms & 5 & 94.4\% \\ 
 3 & 1\,ms & 6 & 98.0\% \\ 
 4 & 1\,ms & 7 & 99.1\% \\
 5 & 2\,ms & 5 & 76.4\% \\
 6 & 2\,ms & 6 & 88.6\% \\
 7 & 2\,ms & 7 & 94.9\% \\
 8 & 2\,ms & 8 & 97.8\% \\[1ex] 
 \hline
\end{tabular}\\
\caption{Eight sets of parameters for method 1. The third column shows the percentage of events that remains after "spikes" removed.}
\label{table:1}
\end{table}

\begin{table}[]
\centering
 \begin{tabular}{||c c c c||} 
 \hline
   & $t_s$ & $n_s$ & percentage \\ [0.5ex] 
 \hline\hline
 1 & 300\,$\mu$s & 4 & 94.0\% \\
 2 & 300\,$\mu$s & 5 & 97.5\% \\
 3 & 300\,$\mu$s & 6 & 98.8\% \\
 4 & 250\,$\mu$s & 3 & 88.0\% \\
 5 & 250\,$\mu$s & 4 & 95.8\% \\
 6 & 250\,$\mu$s & 5 & 98.3\% \\
 7 & 200\,$\mu$s & 3 & 91.4\%\\
 8 & 200\,$\mu$s & 4 & 97.2\% \\
 9 & 200\,$\mu$s & 5 & 98.9\% \\
 10 & 150\,$\mu$s & 3 & 94.3\% \\
 11 & 150\,$\mu$s & 4 & 98.3\% \\
 12 & 150\,$\mu$s & 5 & 99.3\% \\
 13 & 80\,$\mu$s & 3 & 97.6\% \\
 14 & 80\,$\mu$s & 4 & 99.2\% \\
 15 & 80\,$\mu$s & 5 & 99.6\% \\[1ex] 
 \hline
\end{tabular}
\caption{Fifteen sets of parameters for method 2. The third column shows the percentage of events that remains after "spikes" removed.}
\label{table:2}
\end{table}

\begin{figure*}[]
    \centering
    \includegraphics[width=\linewidth]{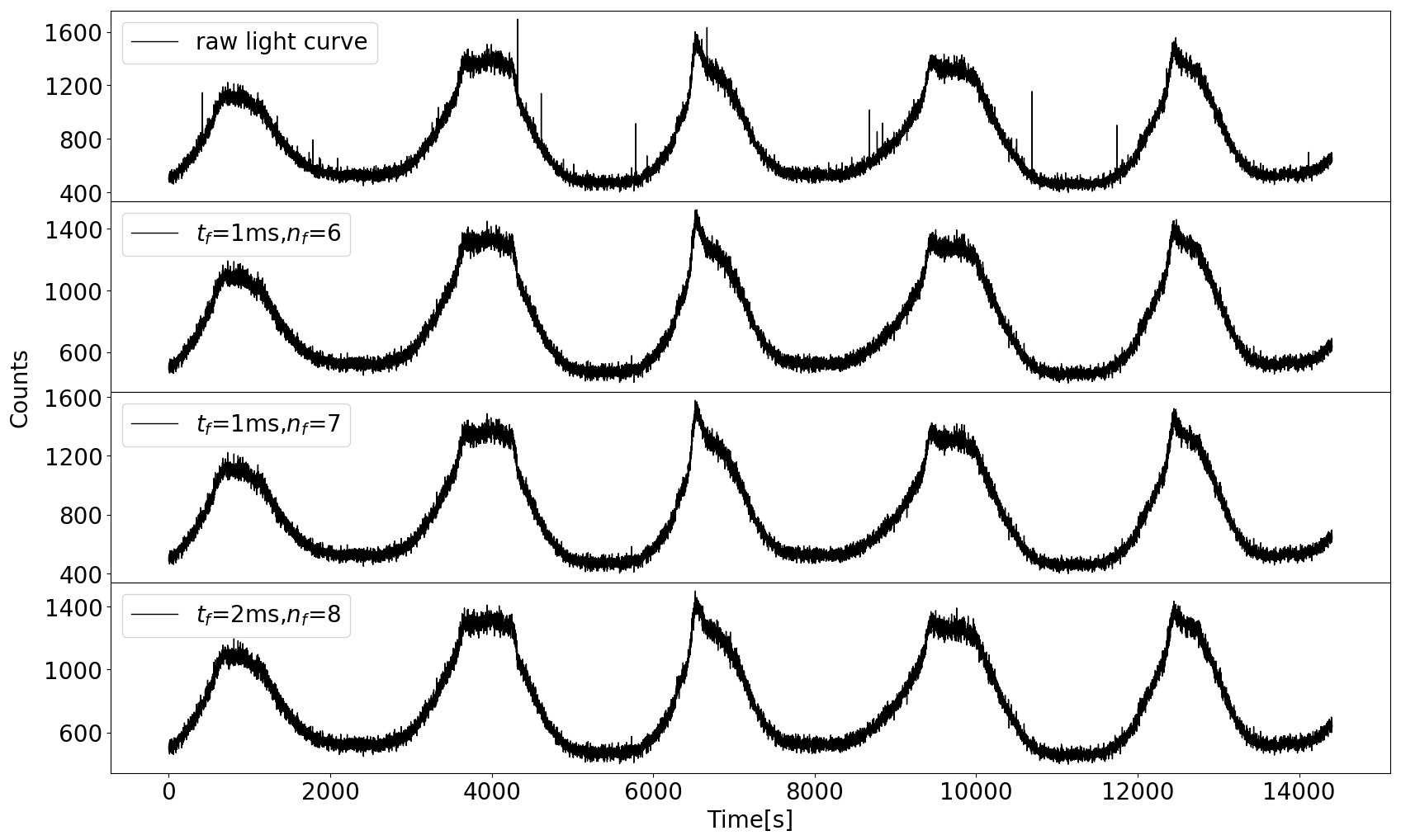}
    \caption{The raw light curve and new light curves generated by method 1 with several sets of parameters, 1\,s binned.}
    \label{FIG:5}
\end{figure*}

\begin{figure*}[]
    \centering
    \includegraphics[width=\linewidth]{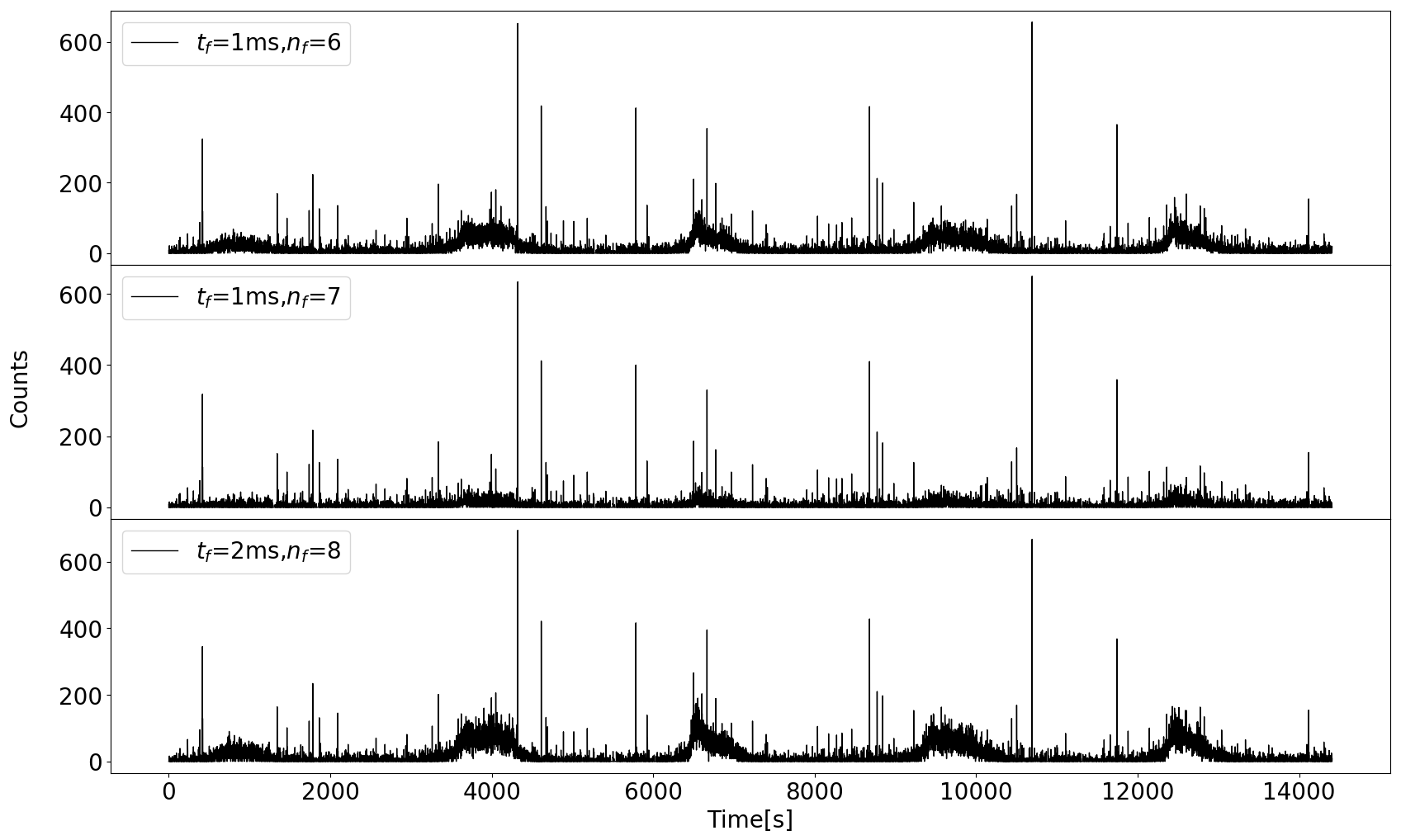}
    \caption{Light curves of spikes using method 1 corresponding to Figure \ref{FIG:5}.}
    \label{FIG:6}
\end{figure*}

We show some of the new light curves that have been removed spike-like events using method 1 in Figure \ref{FIG:5}, and their corresponding “spike” light curves are plotted in figure \ref{FIG:6}. Other light curves with parameters shown in table \ref{table:1} have been omitted since they lost too many normal events. We can see that there are normal events and no sharp spikes on all the new light curves. The percentage of events left is shown in Table \ref{table:1}.\\

\begin{figure*}[]
    \centering
    \includegraphics[width=\linewidth]{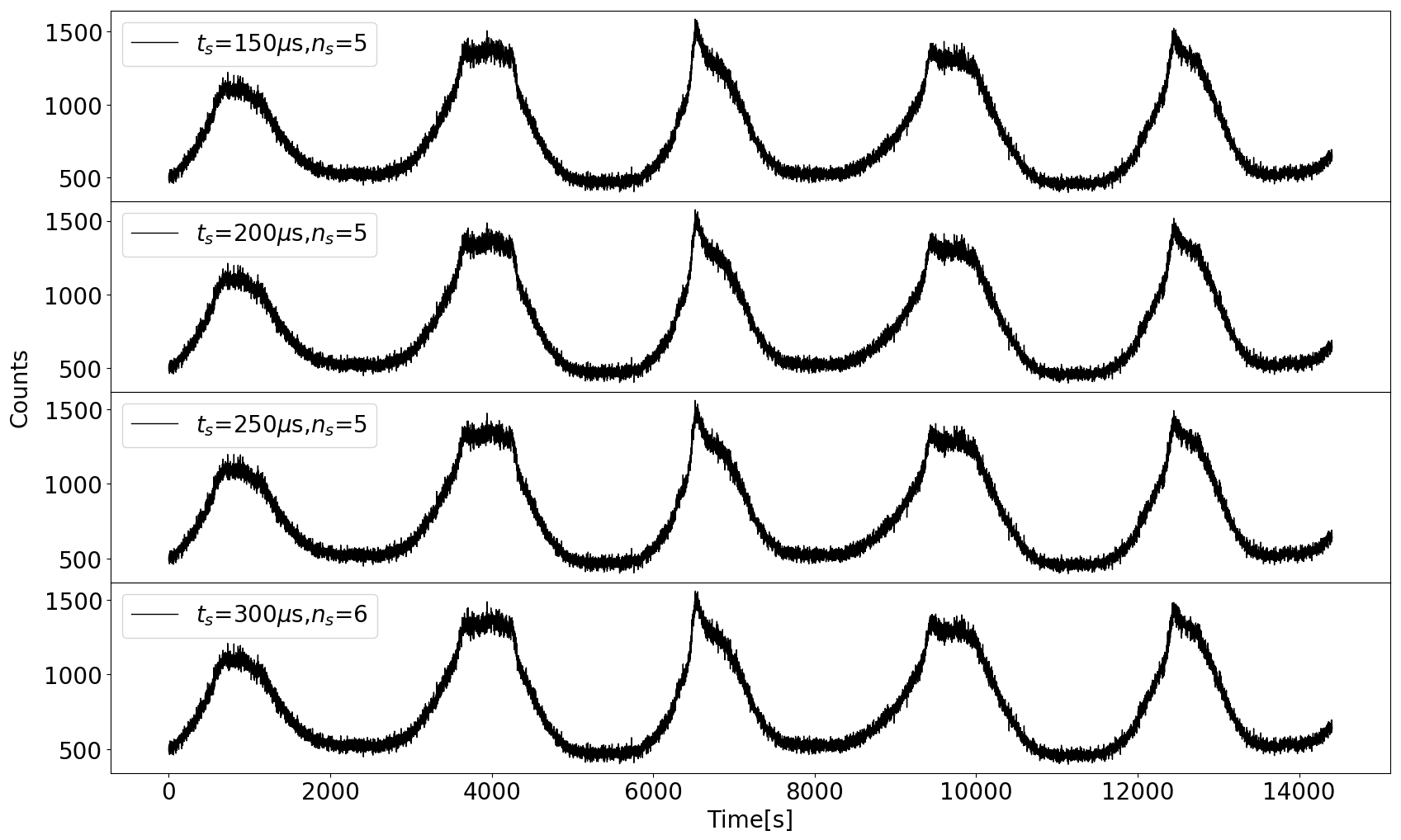}
    \caption{New light curves generated by method 2 with several sets of parameters with time resolution of 1\,s.}
    \label{FIG:7}
\end{figure*}

\begin{figure*}[]
    \centering
    \includegraphics[width=\linewidth]{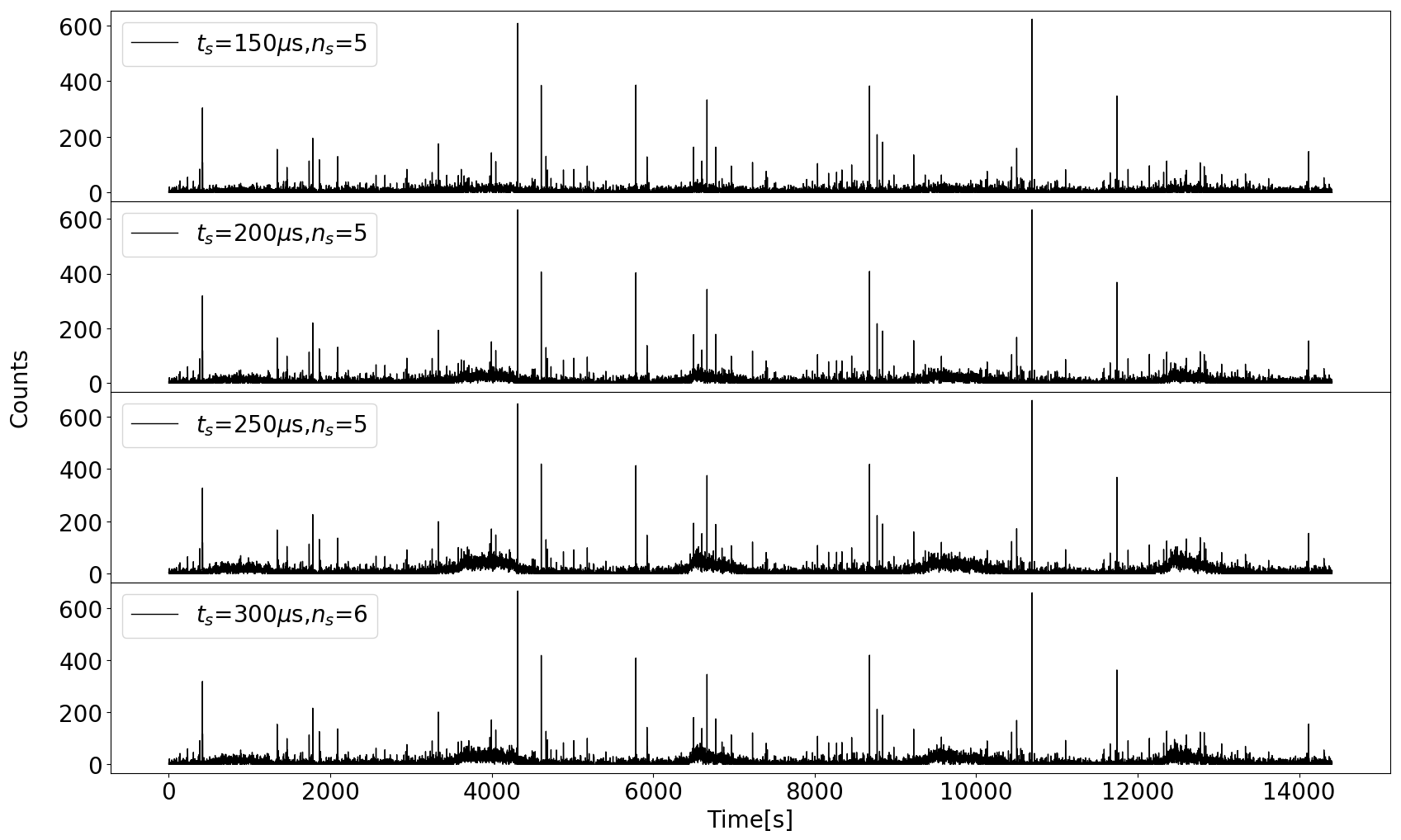}
    \caption{Light curves of spikes using method 2 corresponding to Figure \ref{FIG:7}.}
    \label{FIG:8}
\end{figure*}

Similarly, the light curves of spike-like events removed by using method 2 with relatively good parameters and spike-like events are displayed in Figure \ref{FIG:7} and \ref{FIG:8}, respectively. We use method 2 to remove spikes with these sets of parameters and show them in Figure \ref{FIG:7}. The majority events are remained, and there are hardly spikes on new light curves. The percentage of events left is shown in Table \ref{table:2}.\\

\begin{figure*}
    \centering
    \includegraphics[width=\linewidth]{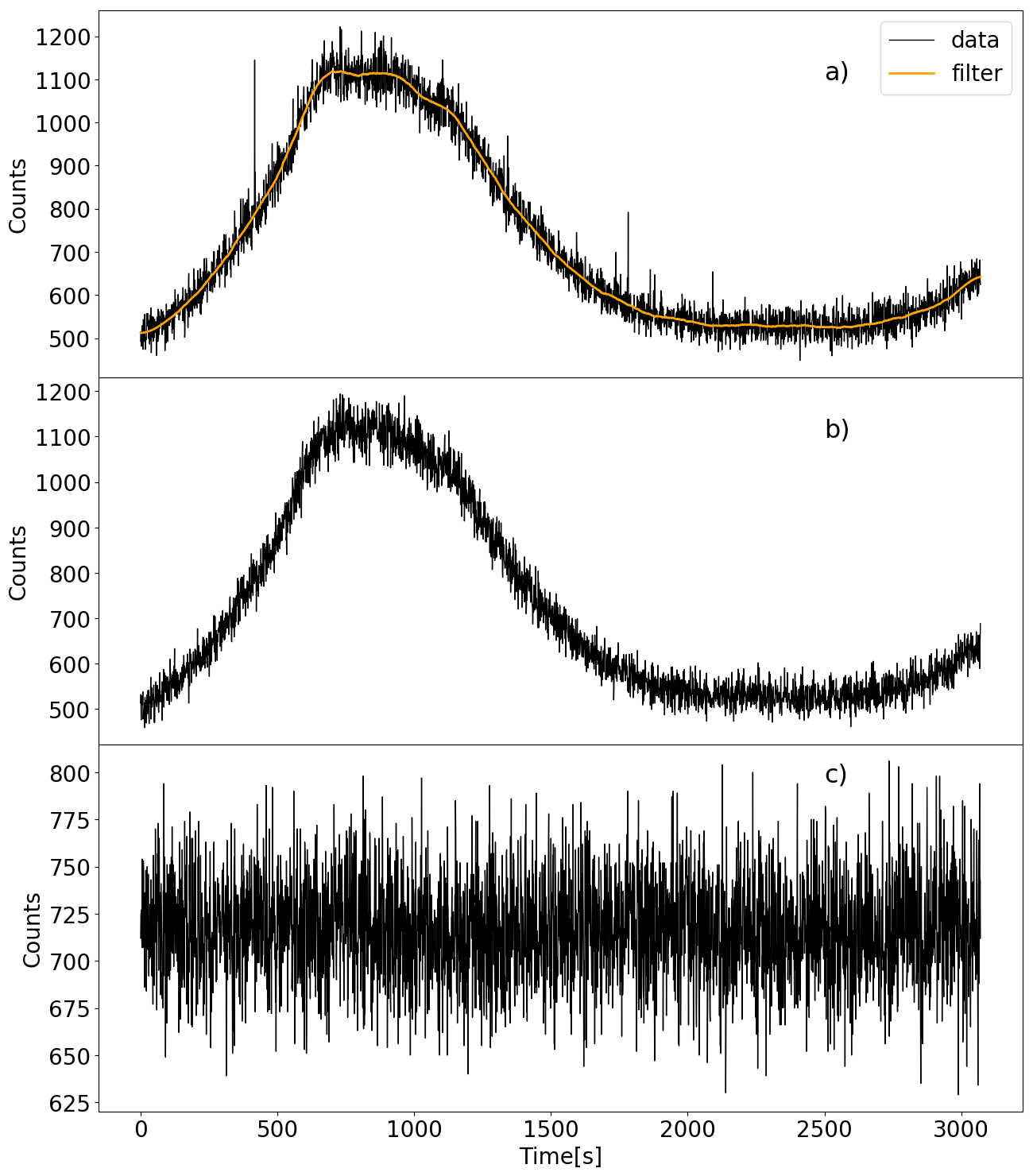}
    \caption{a) The original light curve and its filtered curve. b) The simulated Poisson light curve with filtered expected values. c) A Poisson light curve with a constant expected value.}
    \label{FIG:11}
\end{figure*}

\subsection{The influences of spikes on power spectra}
\label{sec:3.2}
\textcolor{black}{The data we selected for spike analysis can be regarded as white noise dominated at higher frequencies. In addition, HE detectors have different dead times for different types of signal. The  minimum of the dead time is about 3\,$\mu$s for the norm events and 20\,$\mu$s for the large events \cite{2020JHEAp..26...58X}. 
The small dead time will cause powers at frequencies with several thousand Hz a slight decrease. The decrease is relatively tiny and the influence of dead time can be ignored when we focus on the spikes. As a consequence,} we expect a frequency power spectrum without spikes would be distributed as a $\chi^2$ random variable with 2 degrees of freedom\cite{1983ApJ...266..160L} and converge to 2 (Leahy normalization) \cite{1983ApJ...272..256L}. \\

\begin{figure*}
    \centering
    \includegraphics[width=0.9\linewidth]{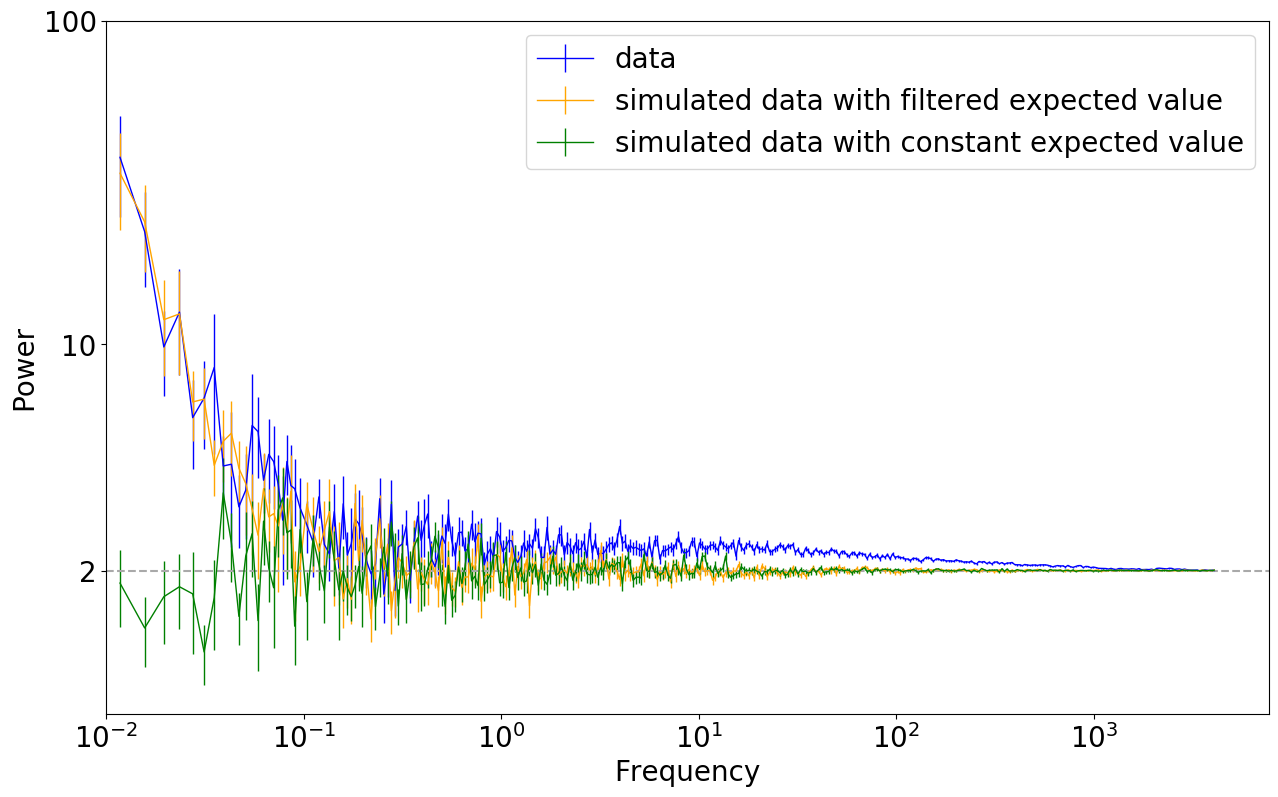}
    \caption{Power spectra of three light curves in Figure \ref{FIG:11}. \textcolor{black}{The blue line, orange line and green line are the Leahy normalized power spectra of the raw data, filtered data and a constant Poisson expected light curve,respectively. The spikes will make the power exceed at more than 1\,Hz. The result of the green line is used to verify the simulation of event list.}}
    \label{FIG:12}
\end{figure*}

In order to find out the influence of spikes to power spectrum, we filter the raw light curve as shown in the panel a) of Figure \ref{FIG:11}. We simulate a Poisson time series and take the counts of filtered curve as the expected values of Poisson function at each second. The new simulated light curve is displayed in the panel b) of Figure \ref{FIG:11} and it has similar shape to the original one without spikes. \textcolor{black}{Then the new light curve is used to simulate a event list by sampling the Poisson distribution. The corresponding power spectrum of the simulated event list in shown in Figure \ref{FIG:12}). Furthermore,} we simulate a Poisson time series using a constant expected value, which is similar to the mean number of events as original light curve as shown in the panel c) of Figure \ref{FIG:11}. \textcolor{black}{This light curve is used to simulate a event list using the same method above to calculate its power spectrum. The Leahy normalized power spectrum must be  converged to 2. As a result, the generation of the simulated event list is verified for the power spectrum generation.}\\

The power spectra of the three light curves as shown in Figure \ref{FIG:11} are generated in Figure \ref{FIG:12}. The power spectrum of simulated data with constant expected value has power value around 2 at all the frequencies as expected. The power spectrum of the simulated data with filtered expected value has the same power-law structure with that of original data at low frequencies, but converges to 2 at high frequencies. Hence, the power-law structure at low frequencies of power spectrum is caused by the overall trend of the raw light curve, while the contribution of the spikes is the additional power between 0.1\,Hz and 300\,Hz which is higher than 2.\\

\begin{figure*}[]
    \centering
    \includegraphics[width=\textwidth]{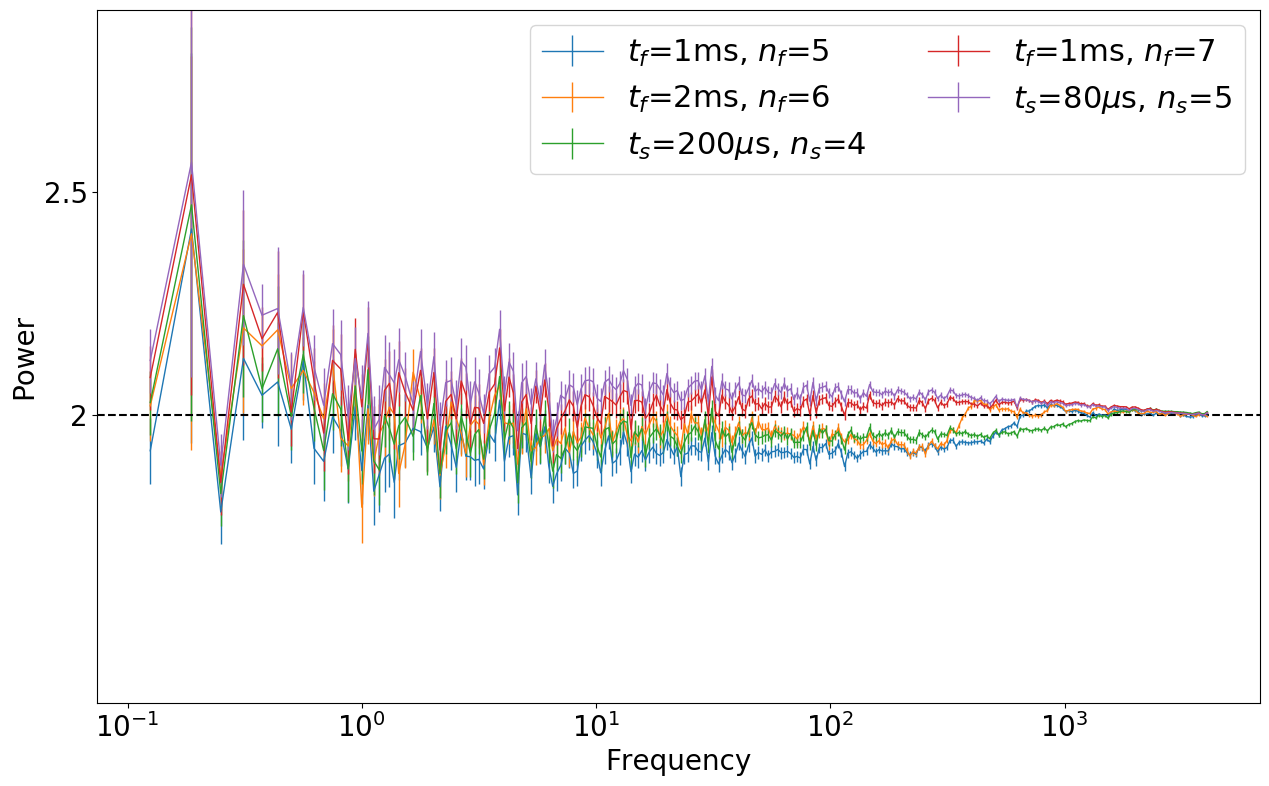}
    \caption{The power spectra that are not suitable for the removal of the spikes. Some power spectra show an average power level under 2, indicating that too many normal events have been removed, such as the blue line($t_f$=1\,ms, $n_f$=5). While some indicate that there are considerable spike events remaining in light curves, leading the power higher than 2, such as the purple line($t_s$=80\,$\mu$s, $n_s$=5).}
    \label{bad_obs}
\end{figure*}

\textcolor{black}{Since we have obtained the expected power spectrum of the light curves without spikes, various parameters of the two methods listed above are applied to screen the raw data and compare the power spectra of the screen data to the expected power spectra. Some power spectra of these parameters deviate a lot from what we expect like the filtered power spectra as shown in Figure \ref{FIG:12}. Some of the power spectra are plotted in Figure \ref{bad_obs} to show why these parameters are discarded. Among these power spectra, some show an average power level under 2, indicating that too many normal events have been removed, e.g. when using method 2 with $t_s$=200\,$\mu$s, $n_s$=4. On the contrary, some power spectra indicate that there are considerable spike events remaining in light curves which make the power higher than 2, e.g. when using method 1 with $t_f$=1\,ms, $n_f$=7.}\\

\begin{figure*}[]
    \centering
    \includegraphics[width=\textwidth]{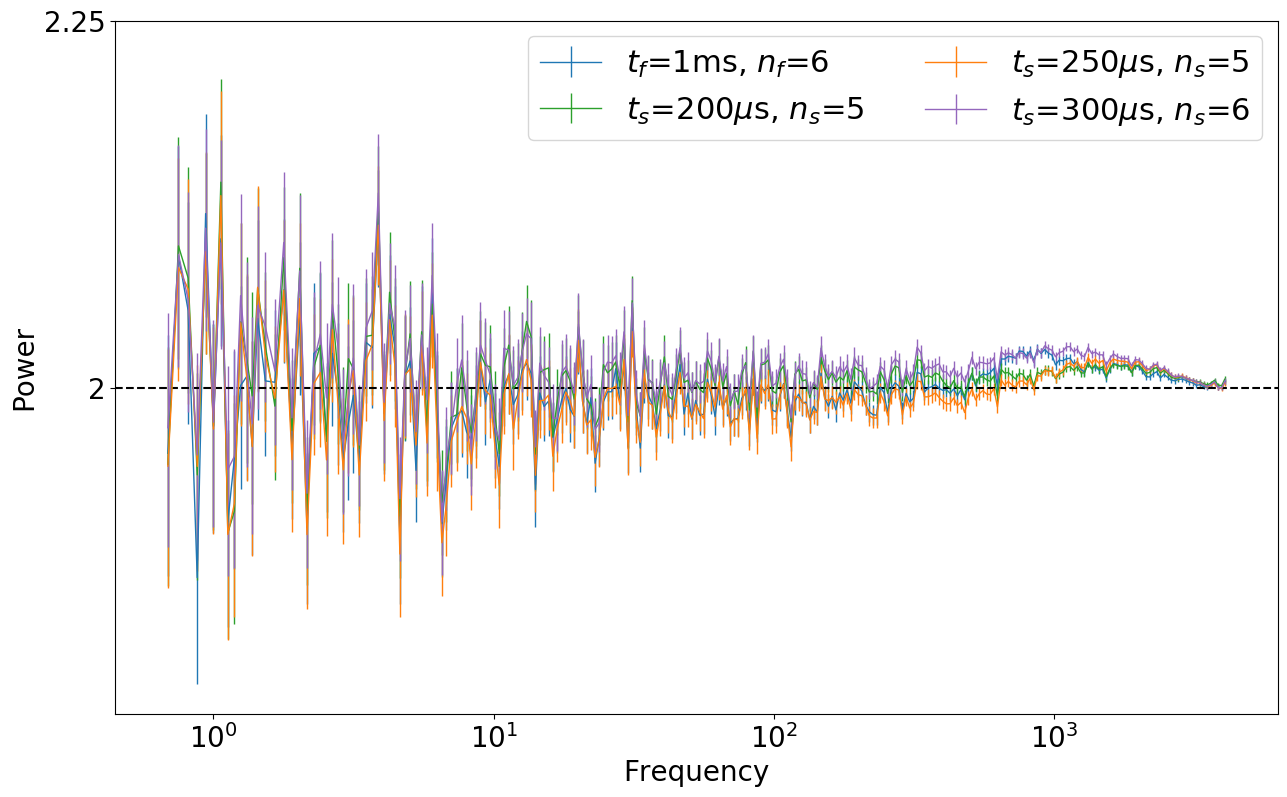}
    \caption{The power spectra that are suitable for the removal of the spikes. However, at high frequencies, their power goes higher and being like bulges, especially when $t_f$=1\,ms, $n_f$=6 and $t_s$=300\,$\mu$s, $n_s$=6. These bulges are still the spikes left in the event list.}
    \label{FIG:obs_good}
\end{figure*}

\textcolor{black}{We select several pairs of parameters of the two methods as candidate to remove spikes.}
Their power spectra are displayed in Figure \ref{FIG:obs_good}. However, we have noticed that at high frequencies, their power goes higher and being like bulges, especially when $t_f$=1\,ms, $n_f$=6 and $t_s$=300\,$\mu$s, $n_s$=6. We think that this bulge is caused by the spikes that are still remained in the event list. According to these power spectra, method 2 with $t_s$=200\,$\mu$s with $n_s$=5 seams to be the best candidate to remove spikes.\\

\subsection{The influences of spikes on energy spectra}
\label{sec:3.3}
\begin{figure*}[]
    \centering
    \includegraphics[width=\linewidth]{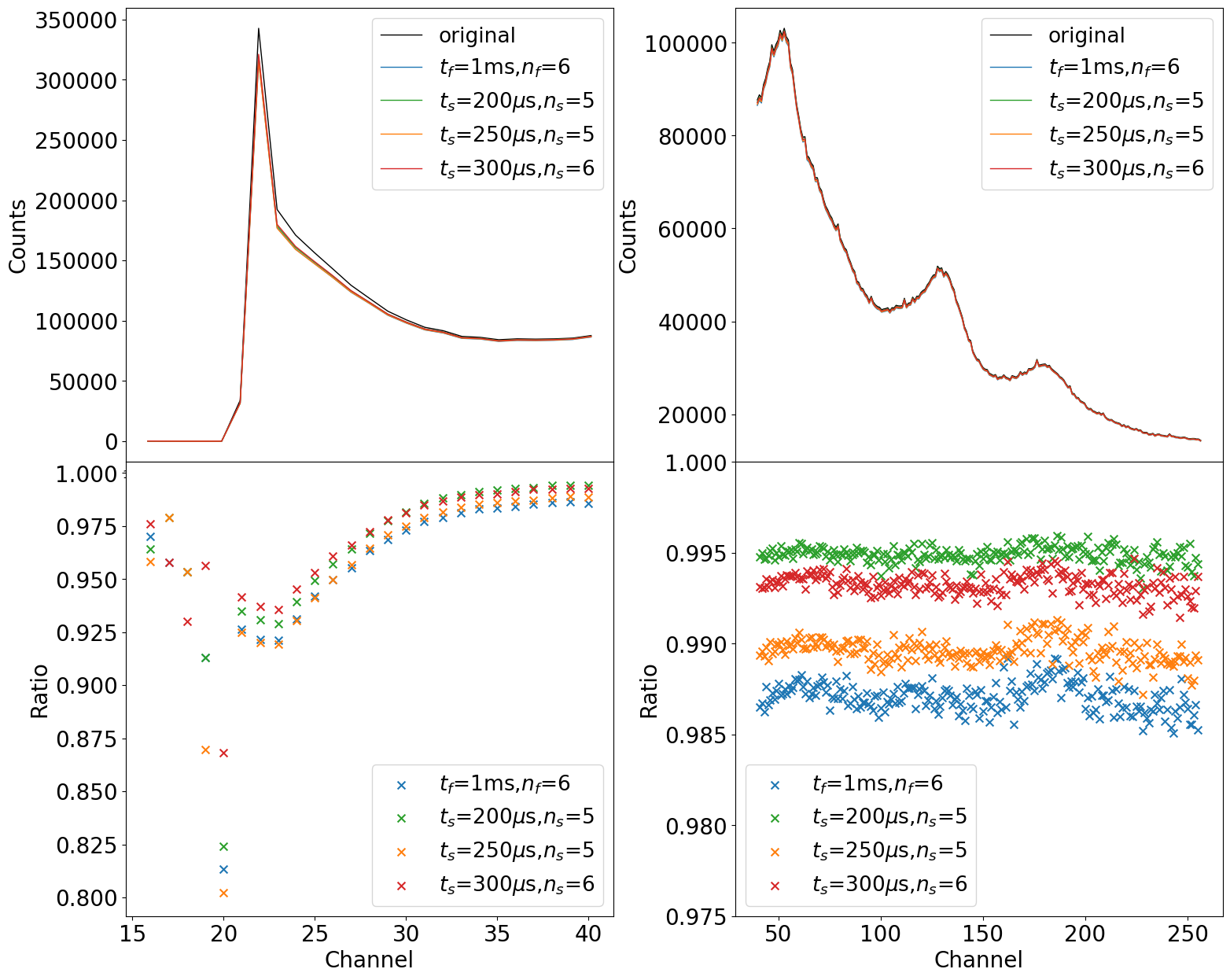}
    \caption{Spectra of new time series whose spike-like events have been removed by method 1 with $t_f=1ms$, $n_f=6$, and method 2 with three sets of parameters, $t_s$=200\,$\mu$s with $n_s$=5, $t_s$=250\,$\mu$s with $n_s$=5, and $t_s$=300\,$\mu$s with $n_s$=6. The below shows ratio between new spectra and the original spectrum. The left are original and new spectra of channels between 15 and 40, the right are the spectra of channels higher than 40.}
    \label{FIG:17}
\end{figure*}

The effects of spikes on energy spectrum are compared only for these new time series whose power spectra are converge to 2 at high frequencies. Therefore, we choose $t_f=1ms$ with $n_f=6$,  $t_s$=200\,$\mu$s with $n_s$=5, $t_s$=250\,$\mu$s with $n_s$=5 and  $t_s$=300\,$\mu$s with $n_s$=6, as candidate to remove spikes. 
The energy spectrum of raw events and spikes removed spectrum with different methods are shown in Figure \ref{FIG:17}. The ratio between each energy channel of the spikes removed spectra and the raw spectra are also displayed in the bottom panel of Figure \ref{FIG:17}. \\

As mentioned in section~\ref{sec:2}, the energy channel of most spike events are under 35 channel. The ratio between original spectrum and the new spectrum should be constant and close to 1 for channels above 40. Among these candidate, the events lost is minimal by using method 2 with $t_s$=200\,$\mu$s, $n_s$=5. Taking accounts of both power spectrum and energy spectrum, the optimal choice for removing the spikes is method 2 ($t_s$=200\,$\mu$s with $n_s$=5).

\captionsetup[figure]{labelfont={color=black}}

\section{The generation and reduction of spikes in HE}
\label{sec:4}
In order to find out the fake triggers in HE, an experiment base on a backup copy of HE detector is taken on ground. 
The HE detector\,(HED) is composed of a NaI(Tl)/CsI(Na) crystal, a PMT R877 and a charge sensitive amplifier. The full width at half maximum\,(FWHM) of a normal NaI(Tl) pulse is about 1.5 $\mu$s, while that of CsI(Na) is 2.5 $\mu$s. \\

\begin{figure*}[]
    \centering
    \includegraphics[width=\linewidth]{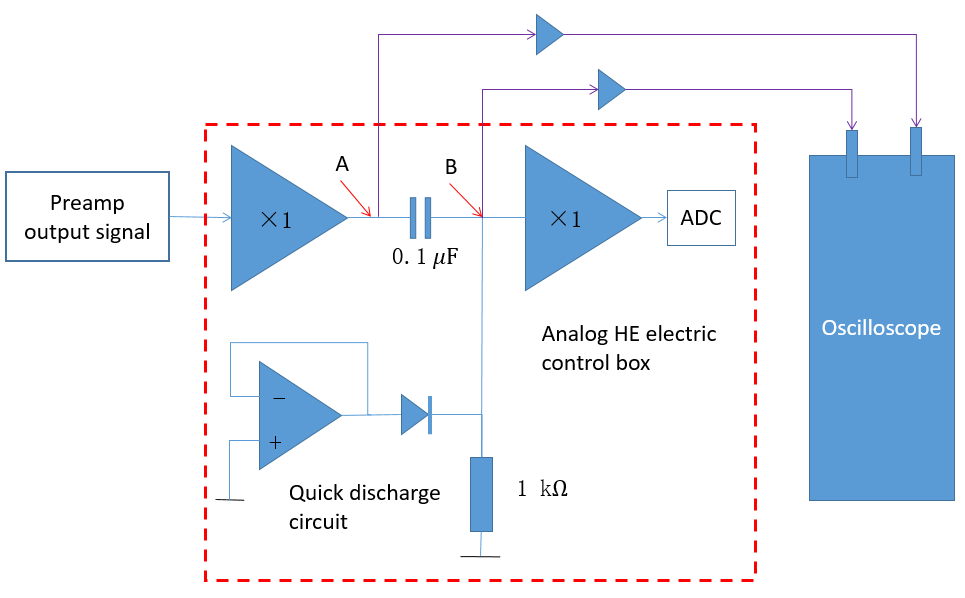}
    \caption{A simplified circuit. There's a blocking capacitor of 0.1\,$\mu$F between point A and B. The oscilloscope is connected to point A and B.}
    \label{FIG:18}
\end{figure*}

Figure \ref{FIG:18} depicts a simplified readout circuit similar to the on-board data acquisition system of the HED. The signals coming from the HED are received by two followed one-unit amplifiers, between which a quick charge circuit is inset. After that the signals are sent to an Analog to Digital Converter\,(ADC). The points A and B are linked with the oscilloscope through two one-unit amplifiers, respectively, so as to measure the pulses in real time. \\

\begin{figure*}[]
    \centering
    \includegraphics[width=0.9\linewidth]{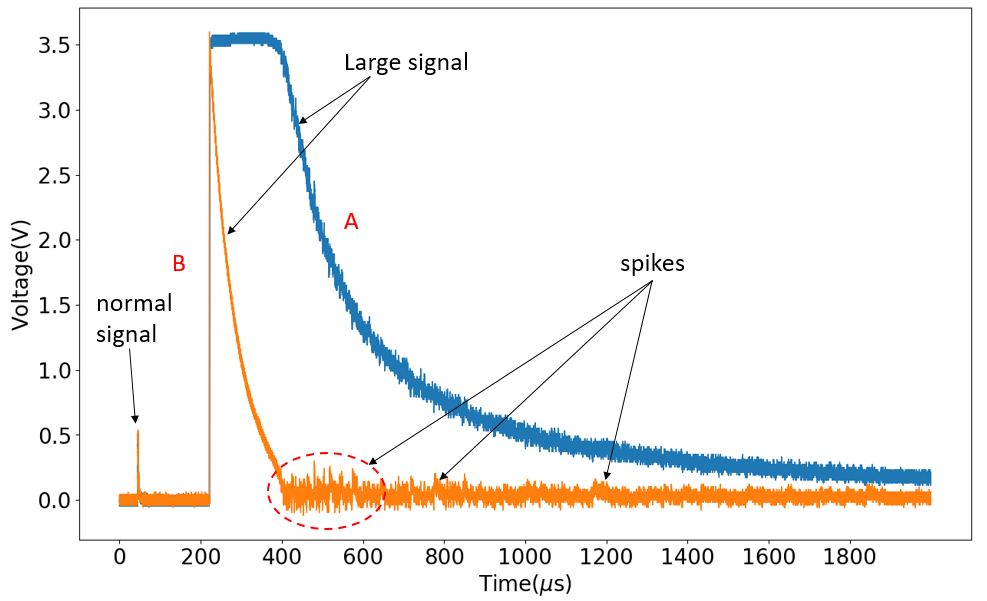}
    \caption{The blue and orange lines represent the big signal from Point A and Point B, respectively. The normal signal and the spikes are also shown by the arrows.}
    \label{FIG:19}
\end{figure*}

We will describe how the big signal generates the spikes in the following.
As displayed in Figure \ref{FIG:19}, after a big pulse in point A, we find that some fake pulses in point B are induced from the fluctuation in steep tail of the big pulse in point A when a high energy charge particle such as a cosmic ray muon deposited large energy in NaI(Tl) or CsI(Na) crystal. So the fake pulses generated in point A can be larger than the threshold for point B and generate the spikes.  \\

In order to study the origin of the spikes, two experiments have been tested. The signal generator is used to take the place of crystal and PMT to generate large signal, the tail of the big pulse is very smooth and no spikes occur. Then we use the LED as the light source to produce the fluorescent photons and illuminate the PMT, the tail of the big pulse will not be smooth, just as shown in the blue line of Figure \ref{FIG:19}. There are some fluctuations on the tail and the spikes appear in this situation. Therefore, the generation of HE spikes is originated from the PMT, and we also have found that if the high voltage of PMT is reduced in flight when HE is working at low gain mode, the spikes in the raw light curve will be very rare. This further proves that the generation of spikes is strongly related to the PMT. \\

\begin{figure*}[]
    \centering
    \includegraphics[width=0.9\linewidth]{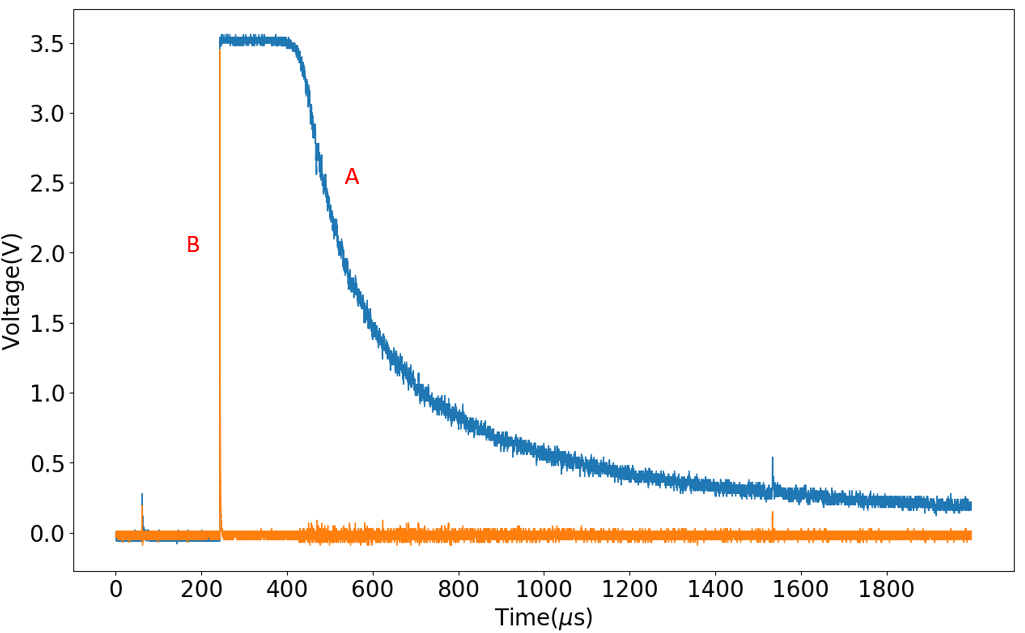}
    \caption{The large signal comes from point A and B and shows in the oscilloscope when the blocking capacitor is changed at 1\,nf. The spikes caused by the large signal can be significantly decreased.}
    \label{FIG:20}
\end{figure*}

Numerical analysis shows that the quick discharge circuit makes a bad contribution. The current charge decay time is 100\,$\mu$s, larger than the fluctuation time in the tail of big pulse. When the discharge circuit works, a down-up fluctuation in point A will become a pulse in point B with a height equal to its peak-to-peak value, almost double the average fluctuation value. In this way, a bigger fluctuation will be amplified and become a fake trigger when its height is above the normal threshold of 100\,mV. We measure those big fluctuations in large signal tail and get an average decay time of about several microseconds. So a capacitance of 1\,nF is chosen to shorten the discharge time. As shown in Figure \ref{FIG:20}, the induced fake pulse amplitude is reduced by more than two thirds, while the normal signal only reduced by one fourth. A smaller capacitance will further reduce signal amplitude and make it pulse width deviate from the design value, meanwhile the fake pulse amplitude will be reduced more. As the fluctuation in the big signal tail is induced by the PMT, it is hard to completely eliminate the spikes. In addition, since the telescope has been on-board, it is impossible to change the circuit, we could only remove spike events through data analysis methods while remaining valid events as many as possible.

\section{Conclusion and discussion}
\label{sec:5}
Directly extracting light curves from raw events files of HE, we have noticed that there are spikes on them. These spikes are some obvious sharp increases on light curve, whose counts are significantly higher than statistical fluctuation. They are not valid x-ray or $\gamma$ events, but results of fake triggers. According to the characteristic of the spikes and spikes' influences on power and energy spectra, we find a best choice ($t_s$=200\,$\mu$s with $n_s$=5) to remove the spikes. \\

According to ground experiments, we also find that the spikes is caused by the large signal in flight which will produce a series of effects after they enter the PMT. These effects make large signals come out from the PMT with long tails which cause electronics continuously triggered and count repeatedly, and finally become spikes on light curve. If we improve the circuit design by changing the blocking capacitor to 1\,nF which would greatly shorten the discharge time and these spikes will also significantly reduced. Since it is impossible to change the circuit when the telescope is on board, the data analysis method is required to remove spikes when producing data products. The method we introduce in this paper has been used in the \insight{} data analysis software.

\section*{ }
\textit{This work is supported by the National Natural Science Foundation of China under grants (No.U1838201, U1838202, U1838105, U1938102, U1938108, U2031205). This work made use of data from the \insight{} mission, a project funded by China National Space Administration (CNSA) and the Chinese Academy of Sciences (CAS).
We also thank the support from the Strategic Priority Program on Space Science, the Chinese Academy of Sciences, Grant No. XDA15020503.}

\section*{Compliance with Ethical Standards}
The authors have no relevant financial or non-financial interests to disclose.

\bibliographystyle{spphys.bst}
\bibliography{ref.bib}

\section*{Affiliations}
\author{Baiyang Wu\textsuperscript{1,2}    \and
        Yifei Zhang\textsuperscript{1}   \and
        Xiaobo Li\textsuperscript{1}     \and
        Haisheng Zhao\textsuperscript{1} \and
        Mingyu Ge\textsuperscript{1}     \and
        Congzhan Liu\textsuperscript{1}  \and
        Liming Song\textsuperscript{1}   \and
        Jinlu Qu\textsuperscript{1}}\\
        
\author{Yifei Zhang}

zhangyf@ihep.ac.cn\\

\author{Xiaobo Li}

lixb@ihep.ac.cn\\

\author{Haisheng Zhao}

zhaohs@ihep.ac.cn\\

\author{Mingyu Ge}

gemy@ihep.ac.cn\\

\author{Congzhan Liu}

liucz@ihep.ac.cn\\

\author{Liming Song}

songlm@ihep.ac.cn\\

\author{Jinlu Qu}

qujl@ihep.ac.cn\\

\begin{itemize}
\item[\textsuperscript{1}] Key Laboratory for Particle Astrophysics, Institute of High Energy Physics, Chinese Academy of Sciences, Beijing 100049, China\\
\item[\textsuperscript{2}] University of Chinese Academy of Sciences, Chinese Academy of Sciences, Beijing 100049, China
\end{itemize}

\end{document}